\documentclass[]{raa}            
\usepackage{graphicx,times}
\usepackage{natbib}

\providecommand{\bm}[1]{\mbox{\boldmath$#1$\unboldmath}}

\begin{document}

   \title{Possible $\gamma$-ray emission of radio intermediate AGN III Zw 2 and its implication on the evolution of jets in AGNs
}

 \volnopage{ {\bf 2010} Vol.\ {\bf 10} No. {\bf 8}, 707--712}
   \setcounter{page}{1}

   \author{Liang Chen
      \inst{1,2,3}
   \and Jin-Ming Bai
      \inst{1,2}
   \and Jin Zhang
      \inst{4}
   \and Hong-Tao Liu
      \inst{1,2}
   }

   \institute{National Astronomical Observatories/Yunnan Observatory, Chinese Academy of Sciences, Kunming 650011, China. {\it chenliangew@hotmail.com; baijinming@ynao.ac.cn}\\
        \and
Key Laboratory for the Structure and Evolution of Celestial Objects, Chinese Academy of Sciences, Kunming 650011, China.
        \and
The Graduate School of Chinese Academy of Sciences, Beijing 100049, China.
        \and
College of Physics and Electronic Engineering, Guangxi Teachers Education University, Nanning, Guangxi 530001, China.
\vs \no
   {\small Received [2010] [May] [10]; accepted [2010] [June] [11] }
}

\abstract{AGNs with hard $\gamma$-ray emission identified so far are radio-loud. III Zw 2 is a radio intermediate AGN with a relativistic jet. We study its spectral energy distribution (SED) and find that the broad band emissions are dominated by the non-thermal emissions from the jet. We model its SED through a synchrotron $+$ inverse Compton (IC) model. The results show that the IC component of III Zw 2 peaks at a few MeV, and the flux density drops rapidly at higher energy with photon index $\Gamma\approx3.3$ above 0.1 GeV. The predicted flux is slightly over the sensitivity of $Fermi$/LAT, but it is not included in the first $Fermi$/LAT AGN catalog. The reason for this may be: 1) that the IC peak is low and the spectrum is very steep above 0.1 GeV, 2) that III Zw 2 is in the low state during the period of the $Fermi$/LAT operation. We also find that III Zw 2 follows similar jet processes as those in $\gamma$-ray AGNs, e.g., the relation between jet power and radiation power, called the blazar sequence.
We suggest that III Zw 2 may be a young source at an earlier stage of jet activity.
}

   \authorrunning{Chen et al.}            
   \titlerunning{Jet in III Zw 2}  
   \maketitle


%

\section{Introduction}

In optically selected samples, quasars with similar optical properties can have very different radio properties, and form a dichotomy in their radio-loudness distribution (i.e., radio-loud, RL and radio-quiet, RQ; Kellermann et al. 1989). It is believed that RL AGNs have powerful relativistic jets, while RQ AGNs have no or very weak jets. In the following research, more and more AGNs were found with intermediate radio-loudness (e.g., Falcke et al. 1996). The study of Very Large Array (VLA) FIRST Bright quasars (FIRST quasars) showed that the radio-loudness distribution is not bimodal (two bumps: RL vs RQ), but rather continuous (White et al. 2000). In addition, VLBI observations reveal that more and more RQ/radio-intermediate AGNs have jet structures at pc scales (e.g.,
Kukula et al. 1998; Ulvestad et al. 2005; Leipski et al. 2006; and references therein) and some with knots move with apparent superluminal velocity, e.g., III Zw 2 (Brunthaler et al. 2000).

The study of jet properties from RQ to RL can offer important information on our understanding on a jet's formation environments, the growth of jets, and the interactions between jets and their environments. In the unified scheme of RL AGNs (see Urry \& Padovani 1995), blazars have relativistic jets with small viewing angles between their jet axis and the line of sight. Their SEDs show two bumps dominated by jet emission, peaking in the infrared (IR) to X-ray bands and in the gamma-ray band, respectively (e.g., Urry \& Padovani 1995). The lower one is generally believed to be the synchrotron emissions of high energy electrons. The higher one is the inverse Compton (IC) emission by the same electron population (e.g., Maraschi et al. 1992; Bloom \& Marscher 1996; Sikora et al. 1994; Dermer \& Schlickeiser 1993; see also the hadronic model; e.g., Mannheim 1993; Aharonian 2000; Atoyan \& Dermer 2003; M{\"u}cke et al. 2003). Until now, all AGNs identified with hard $\gamma$-ray emission are RL. In this paper, we show that the multi band emissions of radio intermediate AGN III Zw 2 are dominated by non-thermal emissions from a jet. We study its jet properties through modeling the spectra energy distribution (SED) based on the synchrotron $+$ IC model, and compare its jet properties with those in $\gamma$-ray AGNs. We conclude that III Zw 2 may be a young source at an earlier stage of jet activity, and its $\gamma$-ray emission may be detected by $Fermi$/LAT. Throughout the paper, a cosmology with H$_{0}=70$ km s$^{-1}$Mpc$^{-1}$, $\Omega_{m}=0.3$ and $\Omega_{\Lambda}=0.7$ is adopted.


\section{General Properties of III Zw 2}

Arp et al (1968) classified III Zw 2 (PG 0007+106, Mrk 1501, $z=0.089$) as a Seyfert I galaxy. It is also included in the PG quasar sample (Schmidt \& Green 1983). Hutchings \& Campbell (1983) derived that III Zw 2 is hosted in a spiral galaxy. The extended radio emission ($\sim$8 mJy at 1.4 GHz, Unger et al. 1987) is very weak compared to the core emission ($\sim$50 to $\sim$100 mJy at 1.4 GHz, Brunthaler et al. 2005). III Zw 2 is also included in the radio-intermediate sample (Falcke et al. 1996). These properties (except for the last one) are typical characteristics of RQ AGN. On the other hand, III Zw 2 possesses some blazar-like properties. Based on VLBA imaging, Brunthaler et al. (2000) reported superlunminal motion in this object with an apparent velocity in units of light speed $\beta_{app}=1.25\pm0.09$ at 43 GHz. This confirms the prediction that radio-intermediate AGNs hold beaming jets (Miller et al. 1993; Falcke et al. 1996; Wang et al. 2006). III Zw 2 shows violent variations up to 30 fold in radio, more than 10 fold in X-rays, and 5 fold in optical; these variabilities are correlated (Salvi et al. 2002). The radio emission is generally believed to be the synchrotron emission of the jet. The correlated variabilities indicate that the broad band emissions originate from the same electron population within the jet. There are also no significant big blue bumps or soft X-ray excesses, which suggests a relatively low emission from the accretion disk (Salvi et al. 2002).

\begin{figure}[h!!!]
   \centering
   \includegraphics[width=9.0cm, angle=0]{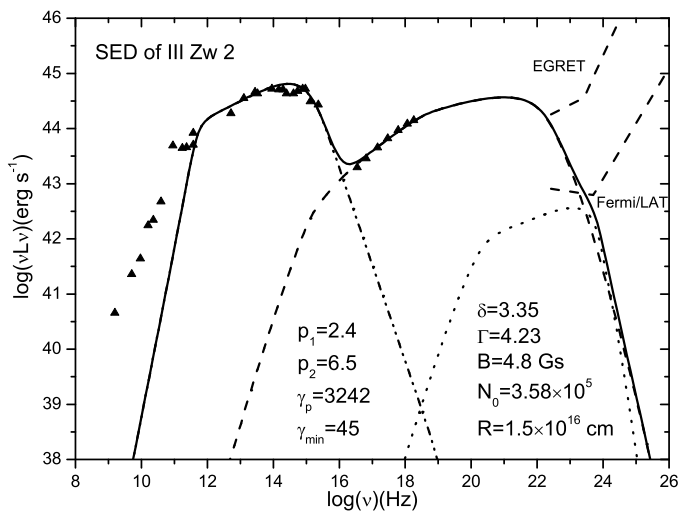}

   \begin{minipage}[]{85mm}

   \caption{Average luminosity from observations are shown as triangles (Salvi et al. 2002). The calculated total SED of III Zw 2 is presented in solid lines. The dotted line is for EC emissions. The dashed line indicates SSC emissions. The dot-dashed line presents the synchrotron emissions. The dashed lines represent the sensitivities of EGRET and $Fermi$/LAT, respectively (McEnery et al. 2004).}
   \label{fig.sed}\end{minipage}
   \end{figure}

These properties show that the broad band emissions of III Zw 2 are dominated by emissions from the beaming jet. We collect the broad band SED of the source from previous archives (see, Fig. \ref{fig.sed}). In the next section, we model its SED and obtain the parameters of the jet.

\section{Assumptions and SED Modeling}

A one zone Synchrotron + IC model is often used to describe the SED of balzars, and can give acceptable fitting results (e.g., Tavecchio et al. 1998,2001; Celotti \& Ghisellini 2008; Ghisellini et al. 2010). In this paper, we use this model to calculate the jet emissions of III Zw 2. The emission region is assumed to be a homogeneous sphere with radius $R$ embedded in the magnetic field $B$. We assume the electron energy distribution to be a broken power law,
\begin{equation}
N(\gamma )=\left\{ \begin{array}{ll}
                    N_{0}\gamma ^{-p_1}  &  \mbox{ $\gamma_{min}\leq \gamma \leq \gamma_{p}$} \\
            N_{0}\gamma _p^{p_2-p_1} \gamma ^{-p_2}  &  \mbox{ $\gamma _p<\gamma\leq\gamma_{max}$.}
           \end{array}
       \right.
\end{equation}
A broken power law distribution which describes the electrons could be the result of the balance between cooling and escape (for more discussion see Kardashev 1962; Sikora et al. 1994; Inoue \& Takahara 1996; Kirk et al. 1998; Ghisellini et al. 1998). The parameters in the model include $R$ and $B$, the electron break energy $\gamma_{p}$, the minimum and maximum energy $\gamma_{min}$, $\gamma_{max}$, the normalized density $N_{0}$, the indexes $p_{1}$, $p_{2}$, the beaming factor $\delta=1/\left[\Gamma\left(1-\beta\cos\theta\right)\right]$, the jet Lorentz factor $\Gamma=1/\sqrt{1-\beta^{2}}$, and the spectrum of external seed photons. In our calculation, the synchrotron self absorption and the Klein-Nishina effect in IC scattering are considered (see, Rybicki 1979; Blumenthal \& Gould 1970).

Popovi\'{c} et al. (2003) derived the inclination angle of the emission line disk $\theta\sim12^{\circ}\pm5^{\circ}$ for III Zw 2. If the emission line disk is the extension of the accretion disk and the jet is perpendicular to the accretion disk, the viewing angle of the jet is $\theta\sim12^{\circ}\pm5^{\circ}$. This combination of the apparent superluminal velocities $\beta_{app}=1.25\pm0.09$ (Brunthaler et al. 2000) gives jet Lorentz factor $\Gamma\approx4.23$ and beaming factor $\delta\approx3.35$. Because of weak accretion disk emissions (Salvi et al. 2002), only the seed photons of the synchrotron radiation (synchrotron self Compton, SSC) and the external radiation (external Compton, EC) from the broad line region (BLR) are taken into consideration in our IC calculation. Kaastra \& de Korte (1988) showed the size of the BLR $\sim10^{18}$ cm, and the photon energy density within the BLR $U_{ext}\approx3.8\times10^{-4}$ erg cm$^{-3}$. From VLBA observations and radio light curves (Brunthaler et al. 2000), it can be derived that the location of the emission region at the beginning of the outburst lies within the BLR. So, the external seed photon energy density measured in the jet's comoving frame is $U_{ext}'\approx(17/12)\Gamma^{2}U_{ext}\approx8.9\times10^{-3}$ erg cm$^{-3}$ (Kaastra \& de Korte 1988; Ghisellini \& Madau 1996) and the spectrum approximates a black body spectrum. The radius of the emission region is determined by the minimum variability time scale $R\sim ct_{var}\delta/(1+z)\approx1.5\times10^{16}$ cm (Jang \& Miller 1997). Other parameters can be derived through modeling the SED, which are shown in Fig. 1.

\section{Results and Discussion}

The calculated SED is shown in Fig. 1, in which the triangles are the average luminosity from observations (Salvi et al. 2002). The total calculated SED is represented by the solid line. The dotted line is for EC emissions. The dashed line indicates SSC emissions. The dot-dashed line presents the synchrotron emissions. It can be seen that the radio emissions cannot be fitted with the model, where the synchrotron self-absorption is significant. Other emitting regions are necessary to fit the radio band (see e.g., Tavecchio et al. 1998; Ghisellini et al. 1998). We can see that the IC component peaks at about a few MeV, and the photon index of III Zw 2 at $\sim$ 0.1-1 GeV is $\Gamma\simeq3.3$. The predicted $\gamma$-ray flux of III Zw 2 is slightly higher than the sensitivity of $Fermi$/LAT and lower than that of EGRET (McEnery et al. 2004). However, III Zw 2 is not included in the first $Fermi$/LAT AGN catalog (Abdo et al. 2010). The reason may be that the spectrum is very steep at the low end of the sensitivity. In another aspect, the long term radio variability shows a $\approx5$ years period (Salvi et al. 2002; Li et al. 2010). The correlation shows that the optical (B band) variability leads radio (14.5 GHz) by $\approx$8 months. It is the same electron population contributing to optical and $\gamma$-ray through synchrotron and IC emissions, thus the simultaneous $\gamma$-ray and optical variabilities may be expected. Derived from radio curves, the recent $\gamma$-ray outburst of III Zw 2 may appear at 2008 April. So, during the 11 months of $Fermi$/LAT operation, III Zw 2 has been in its low state. III Zw 2 has a relatively high flux at soft $\gamma$-ray and hard X-ray bands. The future Hard X-ray Modulation Telescope (HXMT) will constrain its IC component well.

  \begin{figure}[h!!!]
   \centering
   \includegraphics[width=9.0cm, angle=0]{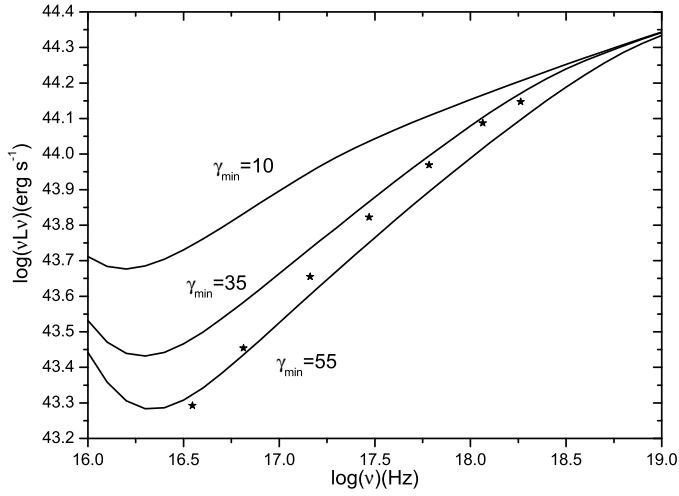}

   \begin{minipage}[]{85mm}

   \caption{Solid lines correspond to X-ray spectra with different electron minimum energies $\gamma_{min}$. It can be seen that this minimum energy may be $\gamma_{min}\approx35-55$.} \end{minipage}
   \label{fig.gammamin}
   \end{figure}

The synchrotron emissions of the electrons approaching the minimum energy are self-absorbed. However, its IC emission is in the X-ray band. This can be used to constrain the minimum electron energy. We give different $\gamma_{min}$ values in SED calculations and find that the best modeling gives $\gamma_{min}\approx35-55$ (see, Fig. 2). Assuming one electron one proton, the jet power can be derived as $L_{jet}=\pi R^{2}\beta\Gamma^{2}cU_{tot}'$ (Celotti \& Fabian 1993; Celotti \& Ghisellini 2008), where $U_{tot}'$ is the energy density in the jet's comoving frame including electrons, protons and magnetic field. The jet radiation power can be calculated by $L_{r}=L_{sed}(4\Gamma^{2}-1)/(3\delta^{4})$ (Celotti \& Ghisellini 2008; Ghisellini 2000). The relation between jet and radiation power is shown in Fig. 3 (stars), in which dots represent $\gamma$-ray AGNs (Celotti \& Ghisellini 2008). We can see that III Zw 2 follows the same law as in $\gamma$-ray AGNs. Thus, they may share the same energy dissipation mechanism. Fig. 4 shows the blazar sequence (Fossati et al. 1998; Ghisellini et al. 1998; Celotti \& Ghisellini 2008). It can be see that III Zw 2 also follows this sequence (stars).

  \begin{figure}[h!!!]
   \centering
   \includegraphics[width=9.0cm, angle=0]{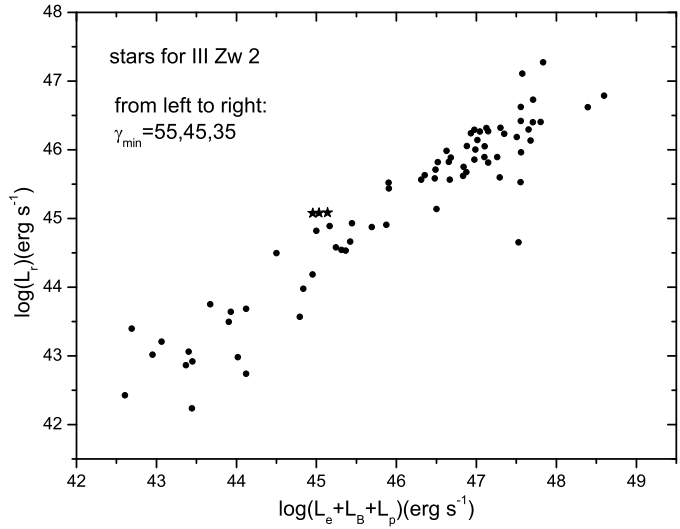}

   \begin{minipage}[]{85mm}

   \caption{Relation between jet power and radiation power. Dots are for $\gamma$-ray AGNs from Celotti \& Ghisellini (2008). Stars are for III Zw 2 from left to right with $\gamma_{min}=55,\ 45, \ 35$.} \end{minipage}
   \label{fig.jetpower}
   \end{figure}

  \begin{figure}[h!!!]
   \centering
   \includegraphics[width=9.0cm, angle=0]{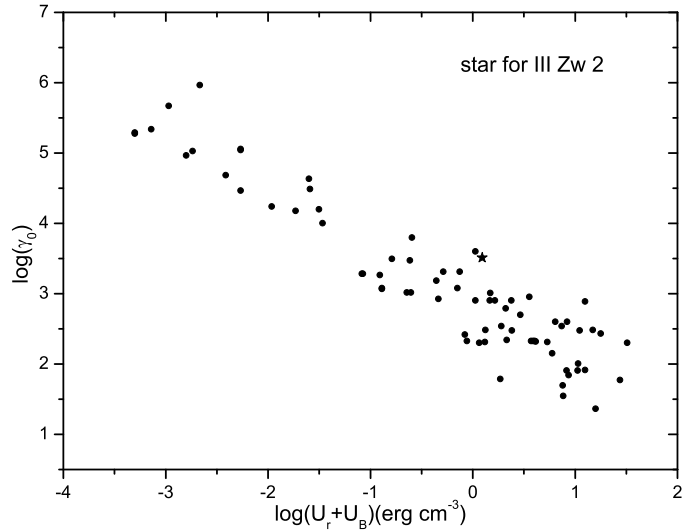}

   \begin{minipage}[]{85mm}

   \caption{Blazar sequence. Dots are from Celotti \& Ghisellini (2008). The star is for III Zw 2.} \end{minipage}
   \label{fig.sequence}
   \end{figure}

From the above analyses, we can see that III Zw 2 has similar jet properties and energy conversion processes as those in $\gamma$-ray AGNs. However, III Zw 2 is a radio intermediate source and other $\gamma$-ray AGNs are RL. There are more evidences that in RL AGNs the radio emission is produced from a more extended region. If an AGN is at the beginning of jet activity, there are not many electrons radiating synchrotron radio emission. Along with jet aging, more electrons accumulate at low energies, which contribute to synchrotron radio emission, and increase the radio-loudness. So, if jet activity is longer enough, the initial non-RL AGN with a relativistic jet will evolve to become RL.

III Zw 2 is radio compact (Brunthaler et al. 2005; Falcke et al. 1999), which is consistent with it being a young source at an early stage of jet activity. In another aspect, for a Compact Steep Spectrum (CSS) and GHz Peaked Spectrum (GPS) source, the radio turnover frequency decreases while the source ages and expands (e.g., O'Dea \& Baum 1997). The expansion of III Zw 2 is parallel to this correlation, which implies that the same physical processes are involved (i.e., synchrotron self-absorption, see Fig. 14. in Brunthaler et al. 2005). After the correction for underestimating of the projected size of III Zw 2, they nearly follow the same correlation, but III Zw 2 lies at the end of the correlation (Brunthaler et al. 2005). This implies that III Zw 2 is the youngest object.

AGNs at an earlier stage of powerful jet activity may form a special subclass with distinctive properties similar to III Zw 2. They may show the compact core, be relatively weak in radio, and have properties associated with relativistic jets, such as broadband non-thermal emission, rapid variability, possible gamma-ray emission, and sometimes apparent superluminal motion. More AGNs at this transition state should be studied to help understand the jet formation environments, the growth of jets and the initial interactions between jets and their environments. Falcke et al. (1996) suggest that radio-intermediate AGNs are radio-weak with relativistic boosting. III Zw 2 is also classified into this class. Some, if not all, of radio-intermediate AGNs may belong to this transition state. The relation between turnover frequencies and the source's age shows that some GSS/GPS sources are young sources. Comparison with III Zw 2 indicates that they may be at a further stage of evolution in their jet activity, which is consistent with their larger radio-loudness. It also suggests that GPS quasars may be special blazars with relativistic jets pointed toward the observer (Bai \& Lee 2005; Bai 2005).

To summarize, the multi-band emissions of III Zw 2 may be dominated by the emissions from the relativistic jet. Its SED can be explained with a synchrotron + IC model, and its possible $\gamma$-ray emission may be detected by further $Fermi$/LAT observations. III Zw 2 may be undergoing a transition state where the AGN is at an earlier stage of jet activity. More studies on such sources are needed to unveil the processes of the jet growth and the initial interactions between jets and their environments.

\normalem
\begin{acknowledgements}
We thank the anonymous referee for helpful comments/suggestions. L.C. thanks the West PhD project of the Training Programme for the Talents of West Light Foundation of the CAS, and National Natural Science Foundation of China (NSFC; Grant 10903025 and 10778702) for financial support. J.M.B. thanks support of NSFC (Grant 10973034) and the 973 Program (Grant 2009CB824800).
\end{acknowledgements}



\label{lastpage}


\begin{thebibliography}{99}
\small \setlength{\itemindent}{-3mm} \setlength{\itemsep}{-0.5mm}
\setlength{\baselineskip}{4.5mm}



\bibitem[Abdo et al.(2010)]{2010ApJ...715..429A} Abdo, A.~A., et al.\ 2010,
\apj, 715, 429


\bibitem[Aharonian(2000)]{2000NewA....5..377A} Aharonian, F.~A.\ 2000, New
Astronomy, 5, 377


\bibitem[Arp(1968)]{1968ApJ...152.1101A} Arp, H.\ 1968, \apj, 152, 1101


\bibitem[Atoyan
\& Dermer(2003)]{2003ApJ...586...79A} Atoyan, A.~M., \& Dermer, C.~D.\ 2003, \apj, 586, 79


\bibitem[Bai
\& Lee(2005)]{2005JKAS...38..125B} Bai, J.~M., \& Lee, M.~G.\ 2005, Journal of Korean Astronomical Society, 38, 125


\bibitem[Bai(2005)]{2005ChJAS...5..207B} Bai, J.-M.\ 2005, Chinese Journal
of Astronomy and Astrophysics Supplement, 5, 207


\bibitem[Bloom
\& Marscher(1996)]{1996ApJ...461..657B} Bloom, S.~D., \& Marscher, A.~P.\ 1996, \apj, 461, 657


\bibitem[Blumenthal
\& Gould(1970)]{1970RvMP...42..237B} Blumenthal, G.~R., \& Gould, R.~J.\ 1970, Reviews of Modern Physics, 42, 237


\bibitem[Brunthaler et
al.(2005)]{2005A&A...435..497B} Brunthaler, A., Falcke, H., Bower, G.~C., Aller, M.~F., Aller, H.~D., \& Ter{\"a}sranta, H.\ 2005, \aap, 435, 497


\bibitem[Brunthaler et
al.(2000)]{2000A&A...357L..45B} Brunthaler, A., et al.\ 2000, \aap, 357, L45


\bibitem[Celotti
\& Fabian(1993)]{1993MNRAS.264..228C} Celotti, A., \& Fabian, A.~C.\ 1993, \mnras, 264, 228


\bibitem[Celotti
\& Ghisellini(2008)]{2008MNRAS.385..283C} Celotti, A., \& Ghisellini, G.\ 2008, \mnras, 385, 283


\bibitem[Dermer
\& Schlickeiser(1993)]{1993ApJ...416..458D} Dermer, C.~D., \& Schlickeiser, R.\ 1993, \apj, 416, 458


\bibitem[Falcke et al.(1999)]{1999ApJ...514L..17F} Falcke, H., et al.\
1999, \apjl, 514, L17


\bibitem[Falcke et al.(1996)]{1996ApJ...471..106F} Falcke, H., Sherwood,
W., \& Patnaik, A.~R.\ 1996, \apj, 471, 106


\bibitem[Fossati et al.(1998)]{1998MNRAS.299..433F} Fossati, G., Maraschi,
L., Celotti, A., Comastri, A., \& Ghisellini, G.\ 1998, \mnras, 299, 433


\bibitem[Ghisellini(2000)]{2000rdgr.conf....5G} Ghisellini, G.\ 2000,
Recent Developments in General Relativity, 5


\bibitem[Ghisellini et al.(1998)]{1998MNRAS.301..451G} Ghisellini, G.,
Celotti, A., Fossati, G., Maraschi, L.,
\& Comastri, A.\ 1998, \mnras, 301, 451


\bibitem[Ghisellini
\& Madau(1996)]{1996MNRAS.280...67G} Ghisellini, G., \& Madau, P.\ 1996, \mnras, 280, 67


\bibitem[Ghisellini et al.(2010)]{2010MNRAS.402..497G} Ghisellini, G.,
Tavecchio, F., Foschini, L., Ghirlanda, G., Maraschi, L.,
\& Celotti, A.\ 2010, \mnras, 402, 497


\bibitem[Hutchings
\& Campbell(1983)]{1983Natur.303..584H} Hutchings, J.~B., \& Campbell, B.\ 1983, \nat, 303, 584


\bibitem[Inoue
\& Takahara(1996)]{1996ApJ...463..555I} Inoue, S., \& Takahara, F.\ 1996, \apj, 463, 555


\bibitem[Jang
\& Miller(1997)]{1997AJ....114..565J} Jang, M., \& Miller, H.~R.\ 1997, \aj, 114, 565


\bibitem[Kaastra
\& de Korte(1988)]{1988A&A...198...16K} Kaastra, J.~S., \& de Korte, P.~A.~J.\ 1988, \aap, 198, 16


\bibitem[Kardashev(1962)]{1962SvA.....6..317K} Kardashev, N.~S.\ 1962,
Soviet Astronomy, 6, 317


\bibitem[Kellermann et al.(1989)]{1989AJ.....98.1195K} Kellermann, K.~I.,
Sramek, R., Schmidt, M., Shaffer, D.~B., \& Green, R.\ 1989, \aj, 98, 1195


\bibitem[Kirk et
al.(1998)]{1998A&A...333..452K} Kirk, J.~G., Rieger, F.~M., \& Mastichiadis, A.\ 1998, \aap, 333, 452


\bibitem[Kukula et al.(1998)]{1998MNRAS.297..366K} Kukula, M.~J., Dunlop,
J.~S., Hughes, D.~H., \& Rawlings, S.\ 1998, \mnras, 297, 366


\bibitem[Leipski et
al.(2006)]{2006A&A...455..161L} Leipski, C., Falcke, H., Bennert, N., {\ H\&uuml}ttemeister, S.\ 2006, \aap, 455, 161


\bibitem[Li et al.(2010)]{2010NewA...15..254L} Li, H.~Z., et al.\ 2010, New
Astronomy, 15, 254


\bibitem[M{\"u}cke et al.(2003)]{2003APh....18..593M} M{\"u}cke, A.,
Protheroe, R.~J., Engel, R., Rachen, J.~P.,
\& Stanev, T.\ 2003, Astroparticle Physics, 18, 593


\bibitem[Mannheim(1993)]{1993A&A...269...67M} Mannheim, K.\ 1993, \aap, 269, 67


\bibitem[Maraschi et al.(1992)]{1992ApJ...397L...5M} Maraschi, L.,
Ghisellini, G., \& Celotti, A.\ 1992, \apjl, 397, L5


\bibitem[McEnery et al.(2004)]{2004ASSL..304..361M} McEnery, J.~E.,
Moskalenko, I.~V.,
\& Ormes, J.~F.\ 2004, Cosmic Gamma-Ray Sources, 304, 361


\bibitem[Miller et al.(1993)]{1993MNRAS.263..425M} Miller, P., Rawlings,
S., \& Saunders, R.\ 1993, \mnras, 263, 425


\bibitem[O'Dea
\& Baum(1997)]{1997AJ....113..148O} O'Dea, C.~P., \& Baum, S.~A.\ 1997, \aj, 113, 148


\bibitem[Popovi{\'c} et al.(2003)]{2003ApJ...599..185P} Popovi{\'c}, L.~{\v
C}., Mediavilla, E.~G., Bon, E., Stani{\'c}, N.,
\& Kubi{\v c}ela, A.\ 2003, \apj, 599, 185


\bibitem[Rybicki
\& Lightman(1979)]{1979rpa..book.....R} Rybicki, G.~B., \& Lightman, A.~P.\ 1979, New York, Wiley-Interscience, 1979.~393 p.,


\bibitem[Salvi et al.(2002)]{2002MNRAS.335..177S} Salvi, N.~J., et al.\
2002, \mnras, 335, 177


\bibitem[Schmidt
\& Green(1983)]{1983ApJ...269..352S} Schmidt, M., \& Green, R.~F.\ 1983, \apj, 269, 352


\bibitem[Sikora et al.(1994)]{1994ApJ...421..153S} Sikora, M., Begelman,
M.~C., \& Rees, M.~J.\ 1994, \apj, 421, 153


\bibitem[Tavecchio et al.(2001)]{2001ApJ...554..725T} Tavecchio, F., et
al.\ 2001, \apj, 554, 725


\bibitem[Tavecchio et al.(1998)]{1998ApJ...509..608T} Tavecchio, F.,
Maraschi, L., \& Ghisellini, G.\ 1998, \apj, 509, 608


\bibitem[Ulvestad et al.(2005)]{2005ApJ...621..123U} Ulvestad, J.~S.,
Antonucci, R.~R.~J., \& Barvainis, R.\ 2005, \apj, 621, 123


\bibitem[Unger et al.(1987)]{1987MNRAS.228..521U} Unger, S.~W., Lawrence,
A., Wilson, A.~S., Elvis, M., \& Wright, A.~E.\ 1987, \mnras, 228, 521


\bibitem[Urry
\& Padovani(1995)]{1995PASP..107..803U} Urry, C.~M., \& Padovani, P.\ 1995, \pasp, 107, 803


\bibitem[Wang et al.(2006)]{2006ApJ...645..856W} Wang, T.-G., Zhou, H.-Y.,
Wang, J.-X., Lu, Y.-J., \& Lu, Y.\ 2006, \apj, 645, 856


\bibitem[White et al.(2000)]{2000ApJS..126..133W} White, R.~L., et al.\
2000, \apjs, 126, 133


\end{thebibliography}
\end{document}